\documentclass[aps,prl,showpacs,twocolumn]{revtex4}
\usepackage{amssymb}
\usepackage{epsf}  %%% 

\begin{document}

\title{On the Enormous Effect of Electric Field on the Crystalline Lattice of the Conductors with Charge-Density Waves.}

\author{V.Ya. Pokrovskii}
\affiliation{Institute of Radioengineering and Electronics of RAS, 125009
Mokhovaya 11-7, Moscow, Russia.}

\begin{abstract}
We discuss deformation of quasi 1-dimensional conductors with charge-density wave (CDW) under the electric field deforming the CDW. In case of ``strong'' CDW-lattice interaction the effect can be treated in terms of the converse piezoelectric effect with enormous piezomodulus, $\sim L_c/\lambda$ times larger than that in ionic crystals ($\lambda$ is the CDW wavelength, $L_c$ is the CDW coherence length, mm scale in the sliding state). The CDW-lattice interaction is likely to be defined by the interband charge transfer (rearrangement of the covalent bonds) with lattice deformation, possible in a number of CDW compounds. The resulting effects, observed or expected, are paving a way towards new-principle actuators, including nanosized ones.
\end{abstract}
\maketitle

At present, one of the promising directions of studies in the field of the charge-density wave (CDW) physics \cite{Grobzor} concerns the elastic \cite{BrillObzor} and dilation \cite{HBZI,PRL,tbp} properties of quasi one-dimensional conductors with CDW. Deformation of the CDW in itself is well studied and its nature is in general understood. It is known, {\it e.g.}, that electric field approaching the threshold value, $E_t$, or exceeding it gives rise to non-uniform CDW deformation (``polarization'')  \cite{INP,braz}. In this case, compression at one contact and strain at the other one spread over the CDW coherence 
length  (more exactly, -- continuity length), $L_c$, which can achieve mm scale for sliding CDW \cite{INP}. More poorly studied are the effects 
of the CDW deformation on the crystalline lattice itself  \cite{munhausen}. Probably, this 
lattice-superlattice coupling was not searched for at first, as it is not 
expected in the simple 1-dimensional model 
\cite{PRL,Mozurkth,Brill1,Mozurkexp,Brill2}. 

Historically, first the CDW-lattice interaction was revealed in the studies of elastic moduli
of the quasi one dimensional conductors with CDW. Decrease of the Young modulus, $Y$, up to 4\%, 
\cite{BrillObzor,Mozurkth,Brill1,Mozurkexp,Brill2} and of the shear modulus, 
$G$, up to 30\% \cite{BrillObzor,BrillShear1,Tas42Ishear} on the CDW depinning under electric field was found. At the same time, such interaction is not a common feature of all the quasi one-dimensional compounds. {\it E.g.}, K$_{0.3}$MoO$_3$, the blue bronze, did not show change of the Young modulus on the CDW depinning: $|\Delta Y/Y| < 5\cdot 10^{-5}$ \cite{NoSoft}.

Different interpretations of the effects have been proposed \cite{BrillObzor}. 
According to \cite{Mozurkth}, if the equilibrium CDW wavelength, $\lambda_{eq}$, normalized by $c$, the lattice period along the chains, for some reason depends on $c$,
a sample deformation draws the CDW into a deformed state. In fact, in the absence of the CDW phase slippage (PS), the pinned CDW changes its period together with the crystal if the latter is deformed -- $\lambda/c={\rm const}$. If at the same time $\lambda_{eq}/c \neq \rm{const}$, the crystal deformation draws the CDW from equilibrium. 
So, pinned CDW contributes to the sample elastic modulus. In the sliding state the internal strains of the CDW rapidly relax, and its elastic contribution drops out \cite{Mozurkth}. 

Consider another case. Let CDW be deformed by an external force. Due to the interaction of the two ``springs'' -- the lattice and the CDW -- the crystal will adjust its dimensions, so that the CDW would approach its equilibrium, and the resulting state will be governed by the minimum of total elastic energy of the host lattice and the CDW \cite{PRL}. This was experimentally observed as metastable length states induced by application of electric field \cite{HBZI} or thermocycling \cite{PRL}; the states, evidently, reveal remnant (metastable) CDW deformation.

As the electric-field induced deformation of the CDW is non-uniform, study of the lattice-CDW interaction from the dilation of the {\it total} crystal turns out to be complicated \cite{PRL,comment}; the basic effect of the electric field on the lattice should be also non-uniform. So, more fruitful, to our understanding, would be search and study of {\it non-uniform} sample expansion under electric field. This idea is also supported by the recently found torsional strain induced in the quasi one-dimensional conductor TaS$_3$ by electric field \cite{tbp}.

In the present Letter we will demonstrate that under electric field exceeding the threshold value one half of the sample should expand, and  the other one -- compress. Experimentally, this effect could be observed, {\it e.g.}, as a shift of the center of a sample with fixed ends. Normalizing this shift by the sample length and dividing by the electric field one will obtain a value roughly $\sim L_c/\lambda$ times above the typical piezomoduli of the known ionic crystals. Numerical estimate for the typical Peierls conductor, TaS$_3$, gives the value $\gtrsim 10^4$\,cm/V, 4--6 orders above the piezomoduli of the known materials. Possible mechanisms of the lattice-CDW interaction are also considered. The most probable explanation, as we suppose, takes into account the possibility of interband charge transfer under crystal deformation. The interaction is expected to be pronounced in compounds with covalent links, say, in trichalcogenides \cite{Meersho}, and may be negligible in the crystals with ionic bonding, like in the blue bronzes.

Let us define the parameter $g$ \cite{PRL,Mozurkth} characterizing the ``strength'' of the CDW-lattice interaction. Let $g+1$ be the coefficient between the relative changes of $\lambda_{eq}$
and $c$ in the case of longitudinal deformation of the crystal under an external force: $\delta \lambda/\lambda=(g+1) \delta c/c$. Clearly, the simple 1-dimensional model gives $g=0$: in this case $\lambda \equiv 2\pi/q = 2/n$, where $n$ is the linear concentration of electrons in the metallic state. As $n$ should change $\propto 1/c$, $\lambda$ will change proportionally to $c$. {\it E.g.}, for a quarter-filled conducting band, $n=1/(2c)$, and $\lambda=4c$. 
Only $g \neq 0$ results in the effects of the CDW-lattice interaction. 
The experimental estimates show that $|g|$ could be of the order of unity and higher \cite{PRL,Mozurkth}. The high value of $g$ also follows from the strain-induced effects in TaS$_3$, which point out the transition of the CDW into the state commensurate with the lattice at a certain value of strain \cite{PT}. We shall discuss the likely origin of the high $g$ value in the end of the paper. Now let us calculate the profile and the value of a sample deformation induced by the electric field taking $g$ as a given value.

For the CDW at the depinning field, the voltage across the sample can be written as

\begin{equation}
\label{V}
V=E_tL+2V_{ps},
\end{equation}
where $L$ is the sample length, $E_t$ is the bulk threshold field, and $V_{ps}$ is the PS voltage, equal to the chemical potential shift at the contacts \cite{Grobzor}. 
This relation means that apart from the force needed to overcome the bulk pinning, the excess field $2V_{ps}/L$ acts upon the CDW chains resulting in the force per unit length $f=(2e/\lambda)\cdot (2V_{ps}/L)$. This force is compensated by the gradient of elastic force $sY_c \frac{1}{q}\frac{{\rm d}^2 \phi}{{\rm d}x^2}$, where $Y_c$ is the CDW elastic modulus, $\phi$ --
its phase gain, and $s$ is the area per conducting chain.
If $L_c>L$, this results in parabolic ``sagging'' of the CDW phase, $\phi(x)$, between the contacts \cite{but}. With $\phi=0$ at the contacts, the maximum
phase gain in the center of the sample is:

\begin{equation}
\label{fi}
\phi(L/2)=\frac{1}{4\pi}eV_{ps}Lq^2/(sY_c),
\end{equation}
and the displacement of the CDW is

\begin{equation}
\label{dxc}
\delta x_c(L/2)=\phi/q=\frac{1}{4\pi}eV_{ps}Lq/(sY_c).
\end{equation}
The longitudinal displacement (deformation) of the sample itself, with respect to the contacts, $\delta x(x)$, follows the ``sagging'' of the CDW with the coefficient $gY_c/Y_L$ \cite{PRL}. By analogy with the converse piezoelectric effect, let us define the ``piezomodulus'' of the CDW conductor, $d_c$, as the displacement of the sample center normalized by the electric voltage applied. 

Two limiting cases worth considering.
The 1st one is a pure short sample, where the volume pinning in the relation~(\ref{V}) can be neglected. At voltages $V \leq 2V_{ps}$ one can substitute  $V/2$ instead of $V_{ps}$ in the relations~(\ref{fi},\ref{dxc}); normalizing $\delta x(L/2)$ by $V$ we obtain:

\begin{equation}
\label{P}
d_c =  \frac{1}{8\pi} \frac{geLq}{sY_L}
= \frac{eg}{4Y_L}\frac{L}{\lambda s}.
\end{equation}
Substituting for TaS$_3$ $g=6$ \cite{PRL}, $\lambda= 12$~\AA, $s=20$~\AA$^2$ \cite{Grobzor}, $Y_L=380$~GPa \cite{BrillObzor}, one obtains for
$L=2$~mm  $d_c=5 \cdot 10^{-6}$~m/V, which by $4 \div 6$ orders exceeds the know values for piezoelectrics.

In the 2nd case -- the more typical one -- one can neglect $V_{ps}$ in the relation~(\ref{V}), and the voltage across the sample is dominated by the value of $E_t$. In this case the ``piezomodulus'' appears $E_tL/2V_{ps}$ times less:

\begin{equation}
\label{PEt}
d_c^{\,'} = \frac{1}{4\pi} \frac{geqV_{ps}}{sY_LE_t}
= \frac{eg}{2Y_L}\frac{V_{ps}}{E_t \lambda s}.
\end{equation}
This value can be, say, $1 \div 2$ order below the above estimate of $d_c$. Still, it is very high in comparison with the piezomoduli of the known materials. In both cases (relations~(\ref{P}) and~(\ref{PEt})) the maximum expected displacement of the middle equals $2d_cV_{ps}$, which for $V_{ps}=3$\,mV \cite{Grobzor} is about 300\,\AA. Such a displacement is quite measurable, {\it e.g.} by means of an atomic-force microscope \cite{skr}.

To understand the physical sense of the enormous $d_c$ for the CDW compounds, we can propose an
estimate for the piezomodulus, $d_i$, of an ion crystal -- see, {\it e.g.}, \cite{enc}:
$d_i^{-1} \sim e/c^2$. In essence, this is an estimate of atomic electric fields. If, for definiteness, we take $\lambda =4c$, the ratio $d_c/d_i$ can be presented as $(Y_0/Y_L)(gL/4\lambda)$, where $Y_0=e^2/(c^2s)$. The physical sense of $Y_0$ is an estimate of the Young modulus (or of the tensile strength) of an ionic crystal \cite{Yionic}. In fact, $e^2/c^2$ is the Coulomb force arising on atomic displacement by $\sim c$ ($c$ is the scale of the extrapolated deformation for which the stress would be equal to $Y_l$). By the way, for $c=3$\,\AA~ we obtain $Y_0=100$\,GPa, in good agreement with the value for TaS$_3$ \cite{BrillObzor}. Thus, the ratio $Y_0/Y_L$ appears to be of the order of unity, and the relation~(\ref{P}) gives the value of $d_c$ roughly $L/\lambda$
times larger than the piezomodulus of a ionic crystals.
The enormous $d_c$ value is associated with the macroscopic
coherence length -- the length of the expansion of the  CDW deformations near the contacts.

In contrast to the conventional piezoelectric effect (and also to the torsional strain in the CDW systems under electric field \cite{tbp}), the longitudinal non-uniform deformation (\ref{P},\ref{PEt}) has no relation to the crystal symmetry point group. The electric field gives rise to a {\it gradient} of the 
crystal strain $\frac{{\rm d}}{{\rm d}x}(\delta c/c)$, but not to the strain $\delta c/c$ in itself. Such phenomenon is known as flexoelectric effect and is observed in liquid crystals, as well as in solids \cite{Tagantsev}. Its observation, in contrast to the piezoelectric effect, does not require absence of the inverse symmetry of the unit cell. 

The calculations above imply a large value
of $g$. To estimate it, one needs to consider the effect of crystal deformation upon the interatomic bonding. $g \neq 0$ implies that the Fermi vector, $k_F$, changes in case of a longitudinal deformation of the crystal. The $\lambda_{eq}/c$ variation could occur due to effects of the strain on the nesting -- the way of the most integrate superposition of the Fermi surfaces, taking in account their curvature \cite{Grobzor}. The nesting determines the $q$-vector value. One can expect growth of curvature of the Fermi surfaces with stretching a specimen of a quasi one-dimensional conductor: the resulting transverse compression would increase the interchain coupling (just to remind, in the absence of interchain coupling, the Fermi surfaces would look as two parallel surfaces). We do not see a way to estimate theoretically the variation of the longitudinal component of the $q$-vector with the sample strain. Neither the sign of the variation is clear. It is likely, that the $q$ deviation from $2k_F$ is quadratic in the curvature of the Fermi surfaces ({\it i.e.} in the transversal band width).

A more plausible explanation of the CDW-lattice interaction, as we think, is related to the transfer of electrons (charge disproportionation) between the valency and the conducting-band states. A feature of the trichalcogenides, MX$_3$, is the variety of the X-X bond lengths, $r_{xx}$ \cite{Meersho}. In TaS$_3$, reducing the S-S length can transform $2S^{2-}$ into $S_{2}^{2-}$ with deliverance of two valency electrons \cite{Meersho}, which cross over the Ta chains, to the conducting band \cite{details}. So, any crystal deformation can result in a change of $k_F$, and, consequently, in a chage of $\lambda_{eq}/c$. 
The strain-induced charge transfer has been discussed also in \cite{BrillObzor} as the mechanism providing the reduction of the elastic moduli on the CDW depinning in NbSe$_3$ and TaS$_3$. 

The above explanation can give $|g| \gtrsim 1$. In fact, a relatively small reduction of the X-X distance, $\delta r_{xx}/r_{xx} <1$, could transfer 2 electrons to the conducting band. Assuming $|\delta r_{xx}/r_{xx}| \sim |\delta c/c|$ we obtain  $|g| \gtrsim 1$. 

As we have noted above, no reduction of the Young modulus has been observed in the blue bronze on the CDW depinning: $|\Delta Y/Y| < 5\cdot 10^{-5}$ \cite{NoSoft}. This could be attributed to the ion bonding in this compound,  in contrast to the trichalcogenides \cite{BrillShear1}. The difference in the type of $Y(E)$ behavior of the blue bronze and of the MX$_3$ compounds makes us trend to the second explanation of the CDW-lattice interaction.

In conclusion, the present Letter states that the crystals of quasi 1-dimensional compounds with CDW can inhomogeneously deform under electric fields about the threshold value for the onset of the CDW sliding. The underlying mechanism of this approximate analogue of the piezoelectric effect is comprised in the CDW elasticity in combination with strain-induced transfer of electrons. The possibility of such transfer is likely to be a feature of the Peierls quasi one-dimensional conductors comprising S or Se \cite{Meersho}. The expected huge non-uniform longitudinal deformation, as well as the recently reported torsional strain \cite{tbp}, open prospects for developing of new-principle electric actuators covering the areas of micro- and nanometer dimensions.

The author is grateful to J.\,W.\,Brill, S.\,N.\,Artemenko, S.\,G.\,Zybtsev and S.\,V.\,Zaitsev-Zotov for useful discussions. The work has been supported by ISTC, RFBR (grant 07-02-91630-MCM\_a). It was performed in the framework of the programs ``New materials and structures'' (RAS, project No 4.21), of the RAS Presidium, and of the CNRS-RAS-RFBR Associated European Laboratory ``Physical properties of coherent electronic states in condensed matter'' including CRTBT and IRE.


\begin{thebibliography}{99}
\bibitem{munhausen}The CDW in itself is a product of the lattice distortion, so the term ``effect of the CDW deformation on the crystal'' is conditional. However, the simple model of two interacting elastic media, as we will see below, contently describes the effects observed.
\bibitem{but}To be exact, the configuration of the CDW deformation is more complex: near the contacts  the dependence $q(x)$ can be exponential \protect\cite{braz}. For simplicity, however, we shall use the model approximation.
\bibitem{details}For TaS$_3$ the pre-known concentration of conducting electrons (1/2 per lattice period per chain) is obtained taking into account that the oxydation state of half the Ta atoms is +4, and of another half -- +5. Atoms of S$^{2-}$ and S$^-$ are distributed in the same proportion \protect\cite{Grobzor,Meersho}. It is natural to assume that the ratio between Ta$^{+4}$ and Ta$^{+5}$ can vary under deformation. Note also, that the isoelectronic compound NbS$_3$ (type II) shows trimerised CDW, which corresponds with 2/3 conducting electrons per lattice period (per chain) \protect\cite{Grobzor}.
\end{thebibliography}
\end{document}